\documentclass[a4paper,12pt]{article}

\usepackage[font=scriptsize]{caption}
\usepackage{lmodern}
\usepackage[T1]{fontenc}

\usepackage[english]{babel}
\usepackage[utf8]{inputenc}

\usepackage{amsmath}
\usepackage{amsfonts}
\usepackage{amssymb}
\usepackage{dsfont}
\usepackage{accents}
\usepackage[mathscr]{eucal}
\usepackage{braket}
\usepackage{bm}
\usepackage{titleref}
\usepackage{varioref}

\usepackage{authblk}

\usepackage{enumitem}

\usepackage[usenames,dvipsnames]{color}

\usepackage{epsfig}
\usepackage{graphicx}

\usepackage{anysize}

\marginsize{3cm}{3cm}{2cm}{2cm}

\title{From physical symmetries to emergent gauge symmetries}

\author[1]{Carlos\ Barcel\'o}
\author[1,2]{Ra{\'u}l\ Carballo-Rubio}
\author[1,3]{Francesco\ Di Filippo}
\author[4,5]{Luis\ J.\ Garay}
\affil[1]{Instituto de Astrofísica de Andalucía (CSIC), Glorieta de la Astronomía, 18008 Granada, Spain}
\affil[2]{Astrophysics, Cosmology \& Gravity Center and Laboratory for Quantum Gravity \& Strings, 
Department of Mathematics \& Applied Mathematics, University of Cape Town, Private Bag, Rondebosch 7701, South Africa}
\affil[3]{Dipartamento di Scienze Fisiche ``E.R. Caianiello''; Universit\`a di Salerno, I-84081 Fisciano (SA) Italy}
\affil[4]{Departamento de Física Teórica II, Universidad Complutense de Madrid, 28040 Madrid, Spain}
\affil[5]{Instituto de Estructura de la Materia (CSIC), Serrano 121, 28006 Madrid, Spain}

\begin{document}

\maketitle

\begin{abstract}
Gauge symmetries indicate redundancies in the description of the relevant degrees of freedom of a given field theory and restrict the nature of observable quantities. One of the problems faced by emergent theories of relativistic fields is to understand how gauge symmetries can show up in systems that contain no trace of these symmetries at a more fundamental level. In this paper we start a systematic study aimed to establish a satisfactory mathematical and physical picture of this issue, dealing first with abelian field theories. We discuss how the trivialization, due to the decoupling and lack of excitation of some degrees of freedom, of the Noether currents associated with physical symmetries leads to emergent gauge symmetries in specific situations. An example of a relativistic field theory of a vector field is worked out in detail in order to make explicit how this mechanism works and to clarify the physics behind it. The interplay of these ideas with well-known results of importance to the emergent gravity program, such as the Weinberg-Witten theorem, are discussed.    
\end{abstract}

\tableofcontents

\section{Introduction}

Despite decades of ongoing research, understanding how to combine the principles of general relativity and quantum mechanics remains elusive \cite{Carlip2001}. One of the stumbling blocks is the importance of gauge symmetries in the mathematical structure of general relativity: as a geometric theory describing the dynamics of spacetime, general relativity cannot be understood without diffeomorphism invariance. There are two possibles stances when it comes to the relative importance given to this idiosyncrasy in research programs aiming to find a theory of quantum gravity. The first possibility is assuming that diffeomorphism invariance is a fundamental property of the next layer, beyond general relativity, in our description of nature. Research programs falling in this category seek for a background independent theory of quantum gravity (see, e.g., \cite{Thiemann2007}). The second possibility is considering that diffeomorphism invariance is an effective feature, so that it emerges together with other relevant features in the low-energy limit of a hypothetical theory which is not necessarily background independent. The ultimate goal of research programs following this hypothesis is finding both an alternative set of principles that would supersede the geometric principles embedded in general relativity and a mechanism ensuring the emergence of the properties of the latter theory.

It is in the latter category, which we call in the following emergent gravity approach,\footnote{This notation is not universal; see, e.g., \cite{Carlip2012} for a discussion on the ramifications of this approach.} in which this work is framed. The singular structure of general relativity makes sensible to expect that this paradigm has chances of being feasible, though the difficulties to overcome cannot be underestimated \cite{Carlip2012}. For instance, the Weinberg-Witten theorem~\cite{Weinberg1980} is usually invoked to argue that this program might be doomed from the start. A related difficulty was recently and sharply pointed out by Marolf~\cite{Marolf2015}. These criticisms are intimately related to the difficulties in understanding how gauge symmetries can emerge in systems with no gauge symmetries~\cite{Carlip2012}, although examples 
of this type of emergence have been described in condensed matter systems (see e.g.~\cite{Baskaran1988}).

These difficulties lie in the different meaning of gauge and physical symmetries. There are various ways to highlight their differences, each of them emphasizing certain peculiarities. In the main body of the text we will make precise statements using the associated Noether currents and charges. However, for the sake of this introductory discussion, probably the most convenient way to stress the different nature of gauge and physical symmetries is the well-known feature that gauge symmetries indicate redundancies in the field-theoretical description of a given system, as it can be made explicit through the occurrence of first-class constraints in its Hamiltonian formulation (see e.g.~\cite{HenneauxTeitelboim1992}). This implies that the fields being used, as well as some combinations of them, cannot be given an operational meaning; that is, do not correspond to observable quantities. Probably one of the most well-know (and debated) examples of this feature is the lack of a meaningful local notion of energy density in general relativity (see e.g.~\cite{MisnerThorneWheeler1974}, Sec 20.4). Hence we might say, in more intuitive terms, that understanding the emergence of gauge symmetries is understanding how and why quantities that are a priori observable (i.e., the fields themselves) become non-observable.

Here we deepen into this question. Working with particular theories has led us to consider a general definition of emergent gauge symmetries that might be applicable to a large class of systems. In the first part of this paper, Sec. \ref{eq:1st}, we present this definition. In order to properly grasp its content, it is necessary to define quite precisely what is meant by local, global, gauge, and physical symmetries. Some of these concepts are often used interchangeably: local and gauge symmetries are sometimes considered synonymous terms, while the same happens with global and physical symmetries. The fine distinctions between these notions do not matter in most situations, though they are of essential importance for our discussion. In Sec. \ref{sec:example} we consider a specific example of physical significance in order to illustrate the meaning of the general definition of emergent gauge symmetries, and check that it leads to sensible results. In this paper we restrict ourselves to the example of an abelian field theory of a relativistic vector field, which is complex enough to illustrate conveniently some of the most relevant features. We discuss several aspects of these systems including the behaviour of observables and how they could emerge form a Galilean underlying structure.
While the definition of emergent gauge symmetries presented here should irrespectively apply to abelian and non-abelian settings, theories falling in the latter cathegory display important peculiarities that are left for future research. Some brief indications of these peculiarities are given in Sec. \ref{sec:dis}, along with a discussion of the relevance of the contents of this paper to the emergent gravity program mentioned in this introduction.

\emph{Notation and conventions:} We work in four spacetime dimensions, though all the results can be equally formulated in arbitrary dimensions. The sign convention chosen for the flat metric $\eta_{ab}$ in Minkowski spacetime is $(-,+,+,+)$. The symbol $\nabla$ denotes always the covariant derivative compatible with the flat metric structure of Minkowski spacetime.

\section{Global or local, physical or gauge?\label{eq:1st}}

\subsection{Some definitions \label{sec:defs}}

Local and global symmetries are classified quite intuitively by the form of their generators. When the generators are independent of the position in spacetime, the symmetry is said to be global (or rigid). If the generators show an explicit dependence on the spacetime coordinates, the symmetry is said to be local.

On the other hand, whether a symmetry of a field theory is gauge or physical is not determined by the form of their generators. In finite-dimensional systems, a typical shortcut to know whether a symmetry is gauge or not might be checking whether their generators depend on functions or on a finite number of parameters~\cite{HenneauxTeitelboim1992,Banados2016}. In field theories one has to be careful with this procedure: a field theory has an infinite number of physical symmetries which are parametrized by functions. Hence the occurrence of local generators cannot be used as a faithful criterion to tell whether a symmetry is gauge or physical (when the generators depend on completely arbitrary functions though, one can be sure that one is facing a gauge symmetry~\cite{Banados2016}). In controversial situations a possible way to proceed is to perform a canonical Hamiltonian analysis, in which gauge symmetries are associated with the occurrence of first-class constraints \cite{Wipf1993}.  An equivalent procedure is to study the associated Noether currents or charges~\cite{HenneauxTeitelboim1992,Banados2016}. 

The evaluation of Noether currents permits to discriminate between gauge and physical symmetries in a simple way. Physical symmetries have associated non-trivial Noether currents that, after integration, lead to non-trivial Noether charges that can be used to parametrize the space of solutions of the classical theory. On the contrary, gauge symmetries lead to trivial (i.e., identically zero) Noether charges. The associated Noether currents $J^a$ take the so-called superpotential form (see, e.g., \cite{Julia1998}):
\begin{equation}
J^a=W^a+S^a,\label{eq:super}
\end{equation}
where $W^a$ is zero on shell, $\left.W^a\right|_{\mathcal{S}}=0$ with $\mathcal{S}$ being the space of solutions, and the superpotential $S^a=\nabla_bN^{ba}$, with $N^{ab}$ being an antisymmetric tensor, is identically conserved.

The classification of gauge and physical symmetries in terms of Noether charges is valid for both finite- and infinite-dimensional systems. Following these definitions, local symmetries are not necessarily gauge symmetries in field theory, as we show in the next section. For the moment let us clarify the definition of physical and gauge symmetries using the well-known example of spinor electrodynamics, with Lagrangian density
\begin{equation}
\mathcal{L}=\mathcal{L}_{\rm EM}+\mathcal{L}_{\rm D}=-\frac{1}{4}F_{ab}F^{ab}+\bar{\psi}\gamma^a(i\nabla_a-A_a)\psi.\label{eq:spielelag}
\end{equation}
This example will help us to fix the language that we will use. Eq. \eqref{eq:spielelag} implies
\begin{equation}
\frac{\delta \mathcal{L}}{\delta \nabla_a\psi}=i\bar{\psi}\gamma^a,\qquad\frac{\delta\mathcal{L}}{\delta A_a}=-\bar{\psi}\gamma^a\psi=-j^a,\qquad \frac{\delta\mathcal{L}}{\delta\nabla_aA_b}=-F^{ab},
\end{equation}
so that the electromagnetic field equations are given by
\begin{equation}
\nabla_aF^{ab}=j^b.\label{eq:spmax}
\end{equation}
The Lagrangian density \eqref{eq:spielelag} is invariant under the local transformations
\begin{equation}
A_a\,\longrightarrow\,A_a+\nabla_a\theta,\qquad \psi\,\longrightarrow\,\exp(-i\theta)\psi,\label{eq:sploctrans}
\end{equation}
with $\theta$ an arbitrary real scalar field, containing as a particular case the global transformations for which $\theta$ is a constant function. The Noether current associated with the global symmetry is given by
\begin{equation}
J^a_{\rm G}=\frac{\delta \mathcal{L}}{\delta \nabla_a\psi}\frac{\delta\psi}{\delta\theta}=\bar{\psi}\gamma^a\psi=j^a,\label{eq:glocurr}
\end{equation}
namely the current that appears on the right-hand side of the Maxwell equations \eqref{eq:spmax}. On the other hand, the Noether current associated with a local symmetry \eqref{eq:sploctrans}, defined using a one-parameter group of transformations $\beta\theta$ with $\beta\in\mathbb{R}$ for a given function $\theta(x)$, is given by
\begin{equation}
J^a_{\theta}=\frac{\delta \mathcal{L}}{\delta \nabla_aA_b}\nabla_b\theta+\frac{\delta \mathcal{L}}{\delta \nabla_a\psi}(-i\theta\psi)=-F^{ab}\nabla_b\theta+\theta j^a.
\end{equation}
This current can be written as
\begin{equation}
J^a_{\theta}=-\theta(\nabla_bF^{ba}-j^a)-\nabla_b(F^{ab}\theta).\label{eq:loccurr}
\end{equation}
This is the superpotential form \eqref{eq:super} expected for gauge symmetries: a term which vanishes on shell plus a superpotential (i.e., an identically conserved quantity). Let us show explicitly that this form implies that the corresponding Noether charges are trivial on shell. The Noether charge associated to the current \eqref{eq:loccurr} is given on shell by
\begin{equation}
\left.Q_{\theta}\right|_{\mathcal{S}}=\int_V\text{d}^3\text{x}\,\nabla_i(F^{i0}\theta)=\int_{\partial V}\text{d}^2\sigma\,n_iF^{i0}\theta.\label{eq:glochar}
\end{equation}
Here $n_i$ is the normal on the two-dimensional boundary surface $\partial V$ of a finite volume $V$ and $\text{d}^2\sigma$ the measure of integration on $\partial V$. Now let us take into account the falloff conditions imposed on the fields at spatial infinity in the limit $\partial V\rightarrow\infty$. The lack of sources of the electromagnetic field at spatial infinity leads to a decay of the electric field $F^{i0}$ of the form $r^{-2}$. Eq. \eqref{eq:glochar} displays a finite value only if $\theta$ behaves asymptotically as $r^{\alpha}$ with $\alpha\leq 0$. Then only the limiting case $\alpha=0$ leads in Eq. \eqref{eq:glochar} to non-trivial charges, but these are the same charges obtained from the Noether current associated with the global symmetries \eqref{eq:glocurr}, corresponding to the electric charge of the system. The remaining cases with $\alpha<0$ lead to identically zero Noether charges. Hence the Noether current \eqref{eq:loccurr} is trivial in the sense that it does not contain any physical information of the system which is not already contained in the Noether current \eqref{eq:glocurr} associated with global transformations.

\subsection{Local does not imply gauge}

Following the definitions in the previous section, local and gauge symmetries are not necessarily identified. Indeed, it is possible to construct a quite illuminating counterexample to this premature identification, using the simple setting of a complex Klein-Gordon field $\phi$. We take this example from the work of Aldaya and collaborators~\cite{Aldaya2010}. Let us consider the Lagrangian density
\begin{equation}
\mathcal{L}_\phi=-{1 \over 2} \nabla_a\phi^{*}\nabla^a\phi- {1 \over 2} m^{2}\left|\phi\right|^{2}.\label{eq:lagphi}
\end{equation}
Being the equations of motion linear on $\phi$, they are invariant under the transformation
\begin{equation}
\phi\rightarrow\phi+\chi\label{eq:shift}
\end{equation}
with $\chi$ itself satisfying the Klein-Gordon equation:
\begin{equation}
(\square-m^{2})\chi=0.
\label{eq:eqforchi}
\end{equation}
On the other hand the Lagrangian density changes in a divergence, thus keeping the action invariant; at first order in $\chi$ one has
\begin{equation}
\delta \mathcal{L}_\phi=\nabla_a B^a,\qquad\qquad
B^a=-\frac{1}{2}\left(\phi^{*}\nabla^{a}\chi+\phi\,\nabla^{a}\chi^*\right).
\end{equation}
The generator of the shift symmetry \eqref{eq:shift} is a functional parameter $\chi$, and hence the symmetry is local (for $m\neq0$ there is not even a global subgroup). However, this symmetry is not gauge: the associated Noether current is non-trivial. This shows that Eq. \eqref{eq:shift} is indeed a physical symmetry, albeit local.

Indeed, taking a specific complex solution $\chi$ of Eq. \eqref{eq:eqforchi}, the Noether current associated with the one-parameter group of symmetries generated by $\beta\chi$ with $\beta\in\mathbb{R}$ is given by
\begin{align}
J^a_{\chi}
=-\frac{1}{2}\left(\chi\,\nabla^a\phi^*+\chi^*\nabla^a\phi-\phi^{*}\nabla^{a}\chi-\phi\,\nabla^{a}\chi^*\right).
\label{eq:noethcurr}
\end{align}
It is straightforward to show that this current is conserved on shell, $\left.\nabla_aJ^a_{\chi}\right|_{\mathcal{S}}=0$. Most importantly, the Noether current \eqref{eq:noethcurr} cannot be written in a superpotential form, which shows that the associated symmetries are physical.

To be more explicit, let us also evaluate the corresponding Noether charges to show that they are non-trivial. The physical meaning of the Noether charges associated with the conserved current $J^a_\chi$ is better grasped if we use the decomposition of $\phi$ in terms of plane waves. Let us consider specific spacetime coordinates $\{t,\mathbf{x}\}$ and a finite box with finite volume $V=L^3$ centered at the origin. Periodic boundary conditions on the edges of the box lead to the decomposition of the solutions of the Klein-Gordon equation in terms of plane waves
\begin{equation}
\phi_{\mathbf{k}}=\frac{e^{i k^ax_a}}{\sqrt{2k^0 V}},\label{20}
\end{equation}
with
\begin{equation}
k^0=\sqrt{m^2+\mathbf{k}^2},\qquad\qquad \mathbf{k}=\frac{2\pi}{L}\mathbf{n}.
\end{equation}
Here the components of $\mathbf{n}$ take integer values, and the value of $k^0$ represents the positive energy solution of the mass-shell condition
\begin{equation}
k_{a}k^{a}+m^{2}=0.
\end{equation}
Hence
\begin{equation}
\phi=\sum_{\mathbf{k}}\left(a_{\mathbf k}\phi_{\mathbf{k}} + b_{\mathbf k}^* \phi_{\mathbf{k}}^* \right)=\sum_{\mathbf{k}}\frac{1}{\sqrt{2k^0V}}\left(a_{\mathbf k} e^{ik^ax_a} + b_{\mathbf k}^* e^{-ik^ax_a} \right).
\label{eq:eqforphi}
\end{equation}
The function $\chi$, being a solution of the Klein-Gordon equation, also displays a decomposition of this form. Let us then consider the symmetry transformation associated with specific plane waves, namely
\begin{equation}
\chi_{\mathbf{k}}^{\rm R}=\phi_{\mathbf{k}}=\frac{e^{i k^ax_a}}{\sqrt{2k^0 V}},
\qquad \qquad
\chi_{\mathbf{k}}^{\rm I}=i\, \phi_{\mathbf{k}}=i\, \frac{e^{i k^ax_a}}{\sqrt{2k^0 V}}.
\label{22}
\end{equation}
The associated Noether charges are then given by
\begin{align}
Q_{\mathbf{k}}^{\rm R}&=\int_V\text{d}^3\text{x}\,J^0_{\chi_{\mathbf{k}}^{\rm R}}=\frac{1}{2}\int_V\text{d}^3\text{x}\left(\chi_{\mathbf{k}}^{\rm R}\,\nabla_0\phi^*+\chi_{\mathbf{k}}^{\rm R*}\nabla_0\phi-\phi^{*}\nabla_0\chi_{\mathbf{k}}^{\rm R}-\phi\,\nabla_0\chi_{\mathbf{k}}^{\rm R*}\right)\nonumber\\
&=\frac{i}{2}(a_{\mathbf{k}}^*-a_{\mathbf{k}}),
\label{Rcharge}
\end{align}
and
\begin{align}
Q_{\mathbf{k}}^{\rm I}&=\int_V\text{d}^3\text{x}\, J^0_{\chi_{\mathbf{k}}^{\rm I}}=\frac{i}{2}\int_V\text{d}^3\text{x}\left(\chi_{\mathbf{k}}^{\rm R} \,\nabla_0\phi^*-\chi_{\mathbf{k}}^{\rm R*}\nabla_0\phi-\phi^{*}\nabla_0\chi_{\mathbf{k}}^{\rm R}+\phi\,\nabla_0\chi_{\mathbf{k}}^{\rm R*}\right)\nonumber\\
&=\frac{1}{2}(a_{\mathbf{k}}^*+a_{\mathbf{k}}).
\end{align}
The charge $Q_{\mathbf{k}}^{\rm R}$ is the conserved charge associated with the invariance of the action under constant shifts of the real part of $a_{\mathbf{k}}$, namely, $(a_{\mathbf{k}}^*+a_{\mathbf{k}})/2$. This is why the resulting value of this charge is precisely the canonically conjugated variable $i(a_{\mathbf{k}}^*-a_{\mathbf{k}})/2$. The same complementarity occurs with the charge $Q_{\mathbf{k}}^{\rm I}$.

Taking instead of Eq. \eqref{22} the solutions
\begin{equation}
\bar{\chi}_{\mathbf{k}}^{\rm R}=\phi^*_{\mathbf{k}}=\frac{e^{-ik^ax_a}}{\sqrt{2k^0 V}},
\qquad \qquad
\bar{\chi}_{\mathbf{k}}^{\rm I}=i \, \phi^*_{\mathbf{k}}=i \, \frac{e^{-ik^ax_a}}{\sqrt{2k^0 V}},
\end{equation}
one gets the conserved charges
\begin{align}
\bar{Q}_{\mathbf{k}}^{\rm R}=\frac{i}{2}(b_{\mathbf {k}}^*-b_{\mathbf {k}}),
\qquad \qquad 
\bar{Q}_{\mathbf{k}}^{\rm I}=\frac{1}{2}(b_{\mathbf {k}}^*+b_{\mathbf {k}}).
\end{align}
Hence not only these symmetries have non-trivial Noether charges, being therefore physical symmetries, but their Noether charges span the entire space of solutions of the theory.
The different solutions of the equation (\ref{eq:eqforchi}) that has to be satisfied by the generator $\chi$ are in one-to-one correspondence to the solutions \eqref{eq:eqforphi} of the Klein-Gordon equation. Hence the set of Noether charges associated with local symmetries $\mathcal{Q}=\{Q_{\mathbf{k}}^{\rm R},\bar{Q}_{\mathbf{k}}^{\rm R},Q_{\mathbf{k}}^{\rm I},\bar{Q}_{\mathbf{k}}^{\rm I}\}$ parametrizes completely the space of solutions $\mathcal{S}$.

The main remark to be taken from this discussion is that local symmetries are not necessarily gauge symmetries, being instead helpful to parametrize the space of classical solutions. This simple observation sets the basis for a mechanism that could give a meaning to the emergence of gauge symmetries, as we explain in general terms in the following discussion.

\subsection{A mechanism for the emergence of gauge symmetries \label{sec:mech}}

Let us consider in general terms a field theory that depends on a set of fields collectively denoted by $\Phi$, with $\mathcal{S}$ the space of solutions of the classical field equations. For our purposes it is only necessary that the system exhibits no gauge symmetries, that is, all the symmetries of the system must be associated with non-trivial Noether charges. In the canonical language, the corresponding Hamiltonian formulation must be free of first-class constraints. In general terms the set $\mathcal{Q}$ of Noether charges associated with all the symmetries (most of them local in character) of the system can be used to parametrize the space of solutions $\mathcal{S}$. Even though this complete set of symmetries exists, in general it cannot be obtained explicitly as finding it would amount to the complete integrability of the system, not possible in general for non-linear systems.

A further necessary condition for our discussion is that some of the degrees of freedom encoded in the fields $\Phi$ are effectively decoupled. That is, there must exist a set of fields $\Psi$ that satisfy a closed system of equations (maybe only approximately), in the sense that non-trivial classical solutions with $\Psi=0$ exist. In other words, the subset $\mathcal{U}\subset\mathcal{S}$ defined by the condition $\Psi=0$ has to be non-trivial, namely $\mathcal{U}\neq\{0\}$ with $\Phi=0$ the trivial solution to the field equations; in general the fields $\Psi$ are not directly a subset of the fields $\Phi$ but specific combinations of them and their derivatives. This permits to define a natural projection
\begin{equation}
\Pi:\mathcal{S}\longrightarrow\mathcal{U}\subset\mathcal{S}.
\end{equation}
The projection $\Pi$ removes the degrees of freedom encoded in $\Psi$, leaving an effective field theory with reduced space of solutions $\mathcal{U}$. 
It is not difficult to come up with general examples that permit to make sense of the existence of this kind of projection associated, for instance, with the consideration of the low-energy regime of a given theory. Indeed, imagine a system with gapless (i.e., massless) excitations and a tower of gapped excitations. Considering the effective description of the system at energies much below the lowest energy gap leads naturally to the projection of any state with momentum $\mathbf{k}$ into the gapless branch of excitations. This removes the massive fields and leaves a theory with massless excitations only.

It is convenient to characterize the projection $\Pi$ in terms of Noether charges. Let us recall the assumption that the set of Noether charges $\mathcal{Q}$ parametrizes completely the space of solutions of the theory, as in the example discussed in the previous section. 
It would be then possible to find a subset $\mathcal{Q}_{\Psi} \subset\mathcal{Q}$ of Noether charges associated with suitable symmetries, most of them of local nature, so that $\Psi=0$ is equivalently given by $\mathcal{Q}_{\Psi}=0$ (this can be thought of as choosing adapted coordinates in configuration space).  There are then two possibilities depending on the interplay between the local symmetries associated with the Noether charges $\mathcal{Q}_{\Psi}$ and the projection $\Pi$:
\begin{enumerate}
\item{There are no local symmetries that preserve the subset $\mathcal{U}$: The effective field theory with space of solutions $\mathcal{U}$ has only symmetries with non-trivial Noether charges, that is, only physical symmetries. The sole effect of the projection $\Pi$ is the elimination of the degrees of freedom parametrized by means of the set of Noether charges $\mathcal{Q}_{\Psi}$.}

\item{There are local symmetries that preserve the subset $\mathcal{U}$: The effective field theory with space of solutions $\mathcal{U}$ has both physical symmetries and local symmetries with trivial Noether charges (or equivalently, trivial currents). These local symmetries are emergent gauge symmetries. Indeed, the effect of the projection $\Pi$ is twofold in this situation. Besides the elimination of the degrees of freedom that stems from its very definition, there is a further reduction on the number of degrees of freedom due to the fact that certain equivalence classes of fields on the subset $\mathcal{U}$ are indistinguishable from an operational perspective.}
\end{enumerate}
This abstract discussion offers a clear framework in which to formulate the emergence of gauge symmetries, following the path of the second case above. Our next aim is to complement this discussion, showing that all the elements and cases that have been introduced can be explicitly constructed in a detailed example of physical relevance. That is, we will discuss in a particular setting the existence and nature of the projection $\Pi$, showing how the structure of the original theory with no gauge symmetries leads either to the first or second cases above. 
A more systematic discussion of the emergence of gauge symmetries along the previous lines could be attained by working in a canonical formalism. The question then will be under which conditions one can introduce in a canonical system first class constraints. We will not pursue this here and concentrate more on the lessons to be obtained from the examples below.

\section{An explicit example \label{sec:example}}

\subsection{Relativistic field theory of a vector field}

The example we will consider is that of a relativistic field theory of a vector field $A^a$. Here $A^a$ has four physical components, and the only symmetry requirement is that of Lorentz invariance. The only remaining condition is that the Lagrangian density is quadratic on the field $A^a$; this will be the most important contribution in any non-linear theory in the limit of small fields. This leads to the Lagrangian density
\begin{equation}
\mathcal{L}_A=-\frac{1}{4}F_{ab}F^{ab}+\frac{\zeta}{2}\left(\nabla_{a}A^{a}\right)^{2}-\frac{m^2}{2}A^a A_a. 
\label{eq:Lwithzetaandm}
\end{equation}
Here $\zeta$ and $m$ are arbitrary parameters. For $\zeta=m=0$ this expression reduces to the standard free electromagnetic Lagrangian density of Maxwell's theory, and hence displays the gauge symmetries that are characteristic of the latter. On the other hand, $\zeta=0$ and $m\neq0$ leads to the Proca action. All of these systems are well known but here we are offering a rather different perspective on their physical content.

The case $\zeta\neq0$, $m=0$ (from now on $\zeta$-theory) is somewhat more subtle. When analyzing the gauge symmetries of Maxwell's theory it is frequent to consider a covariant gauge-fixing condition: the Lorenz condition $\nabla_a A^a=0$. This gauge fixing still leaves some residual gauge symmetry. Residual gauge transformations are given by $A_a \rightarrow A_a+ \nabla_a \chi$ with $\square \chi=0$. The fact that adding the $\zeta$ term to the free electromagnetic Lagrangian density does not break the residual gauge symmetry could lead to think that this symmetry is also gauge in the $\zeta$-theory. However, that is not the case: we show below explicitly that these transformations correspond to physical symmetries for $\zeta\neq0$, using the characterization in terms of Noether currents discussed above. Accordingly, one would expect the $\zeta$-theory to describe four degrees of freedom. This qualitative estimation is consistent with a more rigorous counting based on the Hamiltonian formulation of the theory, which amounts to the determination of the constraints of the theory and their classification into first- and second-class constraints~\cite{HenneauxTeitelboim1992}. Performing this simple analysis we see that no first-class constraints appear in the $\zeta$-theory.

In order to discuss the emergence of gauge symmetries it is convenient to include matter fields to which couple the vector field $A^a$. For simplicity let us consider the Dirac Lagrangian density for fermion fields with the usual coupling to a vector field, namely $\mathcal{L}_{\rm D}$ in Eq. \eqref{eq:spielelag}. Hence the set of fields being considered is $\Phi=\{A^a,\psi\}$. There are two reasons for this particular choice. First, it makes easier to compare the present discussion with the features of the theory of spinor electrodynamics that has been briefly discussed previously in Sec. \ref{sec:defs}. On the other hand, the theory with Lagrangian density $\mathcal{L}=\mathcal{L}_A+\mathcal{L}_{\rm D}$ has been shown to arise as an emergent theory in specific condensed-matter-like systems \cite{Barcelo2014n}, which gives to this example further physical relevance.

The field equations can be obtained straightforwardly from the Lagrangian density $\mathcal{L}=\mathcal{L}_A+\mathcal{L}_{\rm D}$. The field equation for fermion fields takes the form of the standard Dirac equation, while the field equation for the vector field is given by
\begin{equation}
\nabla_bF^{ba}-\zeta\nabla^a\nabla_bA^b-m^2A^a=j^a.\label{eq:elefeq}
\end{equation}
Here $j^a=\bar{\psi}\gamma^a\psi$ is the standard fermion current, which is also the Noether current associated with the global symmetries of the Dirac Lagrangian density $\mathcal{L}_{\rm D}$,
\begin{equation}
\psi\,\longrightarrow\,e^{-i\alpha}\psi,\qquad \alpha\in\mathbb{R}.
\end{equation}
The Lagrangian density \eqref{eq:Lwithzetaandm} is rather well known, particularly for $m=0$ (being in this case the starting point for the Gupta-Bleuler quantization of the electromagnetic field \cite{Itzykson2012}). However, a not-so-well-known fact about it is its invariance under local transformations
\begin{equation}
A_a\,\longrightarrow\,A_a+\nabla_a\chi,\qquad \psi\,\longrightarrow\,e^{-i\chi}\psi,\label{eq:loctrans}
\end{equation}
where $\chi$ is not completely arbitrary but satisfies the equation\footnote{More precisely, any transformation of the form \eqref{eq:loctrans} with $\chi$ satisfying Eq. \eqref{eq:eqforchiwithm} but with a constant source term $\lambda\in\mathbb{R}$ on the right-hand side of the latter would represent a symmetry transformation. Falloff conditions at spatial infinity imply that only the transformations with $\lambda=0$ have a physical meaning.}
\begin{equation}
(\zeta\square + m^2)\chi=0.
\label{eq:eqforchiwithm}
\end{equation}
That these local transformations are indeed symmetries can be shown directly using the field equations \eqref{eq:elefeq}. Despite being local, these do not represent gauge symmetries, but rather physical symmetries. This potential confusion can be settled showing that the associated Noether currents (equivalently, charges) are nontrivial.

For any solution $\chi$ of Eq. \eqref{eq:eqforchiwithm} with we may consider a one-parameter set of symmetries generated by $\beta \chi$ with $\beta\in\mathbb{R}$. Taking into account that the Lagrangian density $\mathcal{L}$ changes in a divergence, the associated Noether current is given by
\begin{align}
J^a_\chi&=\frac{\delta\mathcal{L}}{\delta(\nabla_aA_b)}\frac{\delta A_b}{\delta\beta}+\frac{\delta\mathcal{L}}{\delta(\nabla_a\psi)}\frac{\delta \psi}{\delta\beta}+m^2A^a\chi\nonumber\\
&=-F^{ab}\nabla_b\chi+\zeta\nabla_bA^b\nabla^a\chi+j^a\chi+m^2A^a\chi\nonumber\\
&=-\nabla_b(F^{ab}\chi)+\zeta\nabla_bA^b\nabla^a\chi+\chi(\nabla_bF^{ab}+j^a+m^2A^a).\label{eq:noeth}
\end{align}
The first term in the last identity is a superpotential. On the other hand, the remaining terms are non-zero on-shell:
\begin{equation}
\left.J^a_\chi\right|_{\mathcal{S}}=-\nabla_b(F^{ab}\chi)+\zeta\nabla_bA^b\nabla^a\chi-\zeta\chi\nabla^a\nabla_bA^b.\label{eq:noethonshell0}
\end{equation}
Hence the current \eqref{eq:noeth} leads to non-trivial Noether charges, implying that the local symmetries \eqref{eq:loctrans} are physical symmetries. 
In fact by defining $\varphi=\nabla_aA^a$ one can see that the last two terms in Eq. \eqref{eq:noethonshell0},
\begin{equation}
\zeta \left( \varphi \nabla^a \chi - \chi \nabla^a\varphi\right),\label{eq:noethonshell0b}
\end{equation}
have the same form as the Noether current in the previous example \eqref{eq:noethcurr}, but for a real field instead of a complex field. The analogy with the previous example of the complex scalar field goes further. The theory we are analyzing includes a source-free equation even in the presence of the fermion current $j^a$. The scalar field $\varphi=\nabla_aA^a$ behaves as a free scalar field with no sources: using conservation of $j^a$ one arrives directly to the wave equation
\begin{equation}
(\zeta\square + m^2)\nabla_aA^a= (\zeta\square + m^2)\varphi=0.
\label{eq:eqforphiwithm}
\end{equation}
Eq. \eqref{eq:eqforphiwithm} would be valid for any matter content, given that the vector field that is used on the right-hand side of the electromagnetic field equation \eqref{eq:elefeq} is conserved. Therefore, the local symmetries described by Eqs. \eqref{eq:loctrans}, \eqref{eq:eqforchiwithm} have as Noether charges the Fourier decomposition of the free field equation for $\varphi$. Only for $\zeta=0$ these symmetries correspond to residual gauge symmetries preserving the Lorenz condition, but for $\zeta\neq0$ these are genuine physical symmetries.

Up to now, the example being discussed runs parallel to our discussion in Sec. \ref{sec:mech}. The field theory formulated on the fields $\Phi=\{A^a,\psi\}$ displays global and local symmetries, all of them physical as the study of the associated Noether currents establishes. Furthermore, there is a sector of the theory that decouples dynamically: the field $\varphi=\nabla_a A^a$ follows the evolution equation of a free scalar field. This permits to define the projection $\Pi$ in terms of $\Psi=\{\varphi\}$. The space of solutions $\mathcal{S}$ contains all the solutions to the wave equation \eqref{eq:eqforphiwithm}. However, the lack of sources in this field equation makes natural to consider the projection into the subset $\mathcal{U}$ defined by a trivial value of this scalar field, $\varphi=0$. The action of $\Pi$ is defined on the field $A^a$ (its action over the fermion field $\psi$ is trivial) as
\begin{equation}
\Pi(A^a)=A^a-\nabla^a\xi,
\end{equation}
where $\xi$ is given in terms of a Green function $G(x,x')$ of the d'Alembert operator and the value of $\nabla_aA^a$ as
\begin{equation}
\xi=\int\text{d}^4x'\,G(x,x')\nabla_aA^a(x').\label{eq:xigreen}
\end{equation}
Alternatively, we can use the Helmholtz decomposition of the vector field $A^a$ to write it in terms of its longitudinal and transverse parts as
\begin{equation}
A^a=\bar{A}^a+\nabla^a\xi,\qquad\nabla_a \bar{A}^a=0,\label{eq:hdecomp}
\end{equation}
where $\xi$ satisfies Eq. \eqref{eq:xigreen}. The field content in the vector field $A^a$ can be parametrized in terms of $\bar{A}^a$ and $\xi$ or, equivalently, $\bar{A}^a$ and $\varphi=\nabla_a A^a=\square\xi$. Then the action of $\Pi$ is defined on the pair $( \bar{A}^a,\varphi )$ simply as
\begin{equation}
\Pi[ ( \bar{A}^a,\varphi ) ]= (\bar{A}^a,0 ).\label{eq:pidef}
\end{equation}
Following the parallelism with our previous abstract discussion in Sec. \ref{sec:mech}, the Noether current \eqref{eq:noeth} associated with a local transformation \eqref{eq:loctrans} is rendered trivial on the subset $\mathcal{U}$:
\begin{equation}
\left.J^a_\chi\right|_{\mathcal{U}}=-\nabla_b(\bar{F}^{ab}\chi),\label{eq:noethonshell}
\end{equation}
where $\bar{F}^{ab}=\nabla^a\bar{A}^b-\nabla^b\bar{A}^a$. As in the spinor electrodynamics case discussed in Sec. \ref{sec:defs}, the Noether charges associated with the Noether current \eqref{eq:noethonshell} are trivial. Now, depending on the interplay between the local symmetries \eqref{eq:loctrans} and the subset $\mathcal{U}$ defined by means of the projection \eqref{eq:pidef}, the system would display emergent gauge symmetries or not. Let us consider in detail both cases as discussed in Sec. \ref{sec:mech}.

Before continuing let us mention that Beltr\'an-Jim\'enez and Maroto have analyzed in a series of papers the phenomenological implications of having a non-zero value for $\varphi$ in cosmological scales; see \cite{Jimenez2008,Jimenez2009,Jimenez2009b,Maroto2011} and comments in Sec. \ref{sec:obs}.

\subsection{No emergent gauge symmetries: massive case}

For the massive case $m^2\neq0$, the local symmetries 
\begin{equation}
A_a\,\longrightarrow\,A_a+\nabla_a\chi,\qquad \psi\,\longrightarrow\,e^{-i\chi}\psi,\qquad (\zeta\square+m^2)\chi=0\label{eq:loctransb}
\end{equation}
do not preserve the subset $\mathcal{U}$. Indeed, the last identity in Eq. \eqref{eq:loctransb} implies that any of these local symmetries changes the scalar field $\varphi=\nabla_aA^a$ as (recall that $\zeta\neq0$)
\begin{equation}
\varphi\longrightarrow \varphi-\frac{m^2}{\zeta}\chi.
\end{equation}
Hence the local symmetries \eqref{eq:loctransb} do not preserve the subset defined by $\varphi=0$.

The projection $\Pi$ eliminates one of the four original degrees of freedom that are present for $\zeta\neq0$ and $m\neq0$. The degree of freedom that is removed is parametrized by $\varphi$, and is decoupled from the rest of fields. On the subset $\mathcal{U}$ defined by $\varphi=0$, the resulting theory displays three degrees of freedom and is equivalent to a Proca field theory with $\zeta=0$. Proca's theory yields directly as an equation of motion $\varphi=\nabla_a A^a=0$, being in practical terms absolutely equivalent to the restriction of the theory with $\zeta\neq0$ to the subset $\mathcal{U}$ defined by the condition $\varphi=0$. Proca's theory has therefore 3 degrees of freedom, corresponding in physical terms to a massive photon. Let us also recall that Proca's theory is an example of a theory with second-class constraints, which eliminates directly one degree of freedom.

\subsection{Emergent gauge symmetries: massless case}

For the massless case $m=0$ the situation changes drastically. All the local symmetries \eqref{eq:loctransb} with $m=0$ preserve the subset $\mathcal{U}$. Hence all of these transformations qualify as emergent gauge symmetries, as the associated Noether charges are rendered trivial due to the projection $\Pi$. In other terms it is not possible to operationally distinguish between solutions that fall in the same equivalence class defined by the equivalence relation of being related by the transformations
\begin{equation}
\bar{A}_a\,\longrightarrow\,\bar{A}_a+\nabla_a\chi,\qquad \psi\,\longrightarrow\,\exp(-i\chi)\psi,\qquad \square\chi=0.\label{eq:emesym}
\end{equation}
From the perspective of the fields $\bar{A}^a$ and $\psi$, it is natural to assume that solutions within the same equivalence class represent the same physical solution. As in the massive case discussed above, the projection $\Pi$ removes one degree of freedom from the system through the condition $\varphi=0$. Now the emergent gauge symmetries remove another degree of freedom, leaving only the two degrees of freedom of standard electromagnetism (corresponding to transverse photons). From a canonical perspective, the canonical momentum associated with $A^0$ leads to a first-class (primary) constraint
\begin{equation}
\left.\pi^0\right|_{\mathcal{U}}=\left.\frac{\delta\mathcal{L}}{\delta\nabla_0A^0}\right|_{\mathcal{U}}=\left.\zeta\nabla_aA^a\right|_{\mathcal{U}}=0.
\end{equation}
The resulting theory is then equivalent to the standard electromagnetic theory with both $\zeta=m=0$. Indeed, on the subset $\mathcal{U}$ the electromagnetic field equations \eqref{eq:elefeq} are reduced to the Maxwell equations in the Lorenz gauge  $\nabla_a\bar{A}^a=0$,
\begin{equation}
\square \bar{A}^a=j^a.\label{eq:elefequ}
\end{equation}
Nonetheless, and this is the main point of the paper, these equations arise in a system with no gauge symmetries in its original definition. The emergence of gauge symmetries only make sense under the imposition of certain (natural) restrictions. From this perspective the $\zeta$-theory has 4 degrees of freedom that, in some circumstances, appear effectively as if only two were operating. Hence in the massless case the Lorenz gauge condition would be better understood as a physical condition that ensures the effective decoupling of two of the degrees of freedom in the original theory. In Sec. \ref{sec:dis} we will also discuss this decoupling from the point of view of the stress-energy tensor.

Regarding this example, the $\zeta$-theory, we can finally say that standard electromagnetism with its gauge invariance will appear as soon as Lorentz invariance appears, the mass term is negligible, and the fields satisfy some natural boundary conditions. It is interesting to stress that the claim just made, that electromagnetic \emph{gauge invariance can be seen as deriving from Lorentz invariance}, has a different flavor than the grammatically equivalent assertion made for instance by Weinberg in~\cite{Weinberg1972}, page 289.

\subsection{Some comments regarding quantization}

The discussion and definitions above have been entirely classical. As the example that has been worked out shows explicitly, the definition of the projection $\Pi$ and the corresponding decoupling of certain degrees of freedom seems to rely on conditions that hold only on shell. In quantum field theory some off-shell properties are inevitably explored as a part of the theory. For this reason, one could suspect that this mechanism might break down when introducing quantization. Let us detail why this is not the case. Indeed, it is truly remarkable that the quantum-mechanical translation of the classical description given above is completely parallel to the indefinite, or Gupta-Bleuler quantization of the electromagnetic field (see, e.g., \cite{Itzykson2012} for a textbook discussion). In quantum spinor electrodynamics, some of the features of this particular quantization procedure arise somewhat artificially, such as the introduction of the so-called gauge-breaking term proportional to $(\nabla_a A^a)^2$ (i.e., taking a theory with $\zeta\neq0$), and the subsequent imposition of the Lorenz condition on physical states. 

In the indefinite quantization, a complete set of four independent pairs of creation-annihilation operators are introduced for the vector field operator $\hat{A}^a$. Then, from the canonical commutation relations one obtains the commutation relations for the creation and annihilation operators 
\begin{equation}
[\hat{a}_{A\,\mathbf{k}}, \hat{a}_{B\,\mathbf{k}'}^\dagger]= \eta_{AB}\delta_{\mathbf{k},\mathbf{k}'}.
\label{eq:aCCR}
\end{equation}
We use $A$, $B$ to represent the different polarizations, which being four in this case (corresponding to the four components of $\hat{A}^a$) are formally equivalent to a spacetime index. The negative sign on the left-hand side of \eqref{eq:aCCR} for the $0$-polarization tells us that the Hilbert space will contain states of negative norm. However, these states with negative norm decouple from physical observables, as long as the electromagnetic field interacts with matter fields through a conserved current.

The complete classical decoupling of the additional degrees of freedom of this theory with respect standard electromagnetism occurs when the classical solution $\nabla_a A^a=0$ is singled out by means of the projection $\Pi$. This equation cannot be enforced as an operator-valued equation, though. Remarkably, an equivalent quantum decoupling follows from the (operator-valued) field equations, by requiring the following condition to be satisfied by on-shell quantum states $|\psi\rangle$ in the Hilbert space $\mathcal{H}$:
\begin{equation}
\langle\psi|\nabla_a \hat{A}^a|\psi\rangle=0.\label{eq:quantcond1}
\end{equation}
This condition guarantees that the condition $\nabla_aA^a=0$ is verified in the classical limit. A slightly stronger but equivalent condition takes the form
\begin{equation}
\nabla^a \hat{A}^{(+)}_a |\psi\rangle=0,
\label{eq:constraint}
\end{equation}
where the $(+)$ represents the annihilation part (positive energy) of the operator $\nabla^a \hat{A}_a$. Either \eqref{eq:quantcond1} or \eqref{eq:constraint} would represent the analogue of the projection $\Pi$ in the quantum theory. Let us stress that this is one of the points that work straightforwardly in the abelian case, but which could lead to subtleties for non-abelian interactions.\footnote{A possible route to overcome this problem is starting with abelian settings and then consider the nonlinear completions that extend naturally this linear decoupling.}

\subsection{Emergent gauge symmetries and observables \label{sec:obs}}

We have been discussing how physical symmetries could lead to emergent gauge symmetries under certain natural conditions. The emergence of gauge symmetries implies that some quantities, such as some of the very degrees of freedom encoded in the basic fields $\Phi$, lose their operational meaning. In this section we focus our attention in this feature using the example developed above with values of the parameters in Eq. \eqref{eq:Lwithzetaandm} given by $\zeta\neq0$ and $m=0$.

Leaving aside the most basic observable quantities that lose their operational meaning, namely the very fields of the theory, let us focus on a specific quantity with physical significance: the stress-energy tensor. For the sake of simplicity we intentionally omit matter fields in this discussion (their explicit consideration does not change the conclusions). Application of Hilbert's prescription leads to the expression
\begin{align}
	\label{eq:SETzeta}
T_{ab}&=  -F_{ac}F_{\,\,\,b}^{c} + \eta_{ab}\mathcal{L}_A
\\
& +\zeta A_{a}\nabla_{b}\nabla_{c}A^{c}+\zeta A_{b}\nabla_{a}\nabla_{c}A^{c}
-\zeta\eta_{ab}\nabla_{d}(A^{d}\nabla_{c}A^{c}).
\nonumber
\end{align}
What we want to highlight is that the stress-energy tensor above is not in general invariant under the local symmetries
\begin{equation}
A_a\,\longrightarrow\,A_a+\nabla_a\chi,\qquad \square\chi=0.\label{eq:emesym1}
\end{equation}
Let us recall that $\varphi=\nabla_aA^a$ behaves as a free scalar field, verifying Eq. \eqref{eq:eqforphiwithm}. The most natural solution of this equation is $\varphi=0$, but we could also propose any solution of the form \eqref{20} (working again on a finite volume $V$), namely
\begin{equation}
\varphi_{\mathbf k}=\frac{e^{ik^ax_a}}{\sqrt{2k^0 V}}\label{eq:planewave},
\end{equation}
with $k^a k_a=0$ and $k_0>0$, or any superposition of them (here we understand $\varphi$ as a complex extension of the real field). A given functional form of $\varphi$ can be fixed a priori precisely because $\varphi$ is not changed by the transformations \eqref{eq:emesym1}.

If one selects the solution $\varphi=0$ then the stress-energy tensor in Eq. (\ref{eq:SETzeta}) becomes the standard Maxwell stress-energy tensor (containing only the first two terms in the latter equation). In this case the stress-energy tensor is invariant under the emergent gauge symmetries \eqref{eq:emesym1}. However if we choose a non-zero solution like (\ref{eq:planewave}) the stress-energy tensor happens to be non-invariant. Under a local symmetry \eqref{eq:emesym1} one has
\begin{equation}
T_{ab}\longrightarrow T'_{ab}=T_{ab}+i\zeta\left(k_{a}\nabla_{b}\chi+k_{b}\nabla_{a}\chi-\eta_{ab}k^c\nabla_c\chi\right)\frac{e^{ik^ax_a}}{\sqrt{2k^0 V}}.\label{eq:setvar}
\end{equation}
Note that the stress-energy tensor is defined up to a superpotential, i.e., an identically conserved term. One might think that it could be possible to add a suitable superpotential to the stress-energy tensor \eqref{eq:SETzeta} to make the overall expresion invariant under the local tranformations \eqref{eq:emesym1}. But this is indeed not possible. Given that adding a superpotential does not change the conserved charges, the simplest way to show this is to prove that the associated conserved charges $P_a$ are not invariant, where
\begin{equation}
P_{a}=\int_V\text{d}^{3}\text{x}\,T_{a0}.
\end{equation}
From Eq. \eqref{eq:setvar}, their finite change $\Delta P_a=P'_a-P_a$ under a local symmetry \eqref{eq:emesym1} is given by
\begin{equation}
\Delta P_{a}=i\zeta\int_V\text{d}^{3}\text{x}\left(k_{a}\nabla_{0}\chi+k_{0}\nabla_{a}\chi-\eta_{a0}k^c\nabla_c\chi\right)\frac{e^{ik^dx_d}}{\sqrt{2k^0 V}}.
\end{equation}
To show that the charges are not invariant it is enough to show that this is valid for a particular generator of the local symmetry \eqref{eq:emesym1}. Let us then make the choice
\begin{equation}
\chi_{\tilde{\mathbf{k}}}
=\varphi^*_{\tilde{\mathbf{k}}}
=\frac{e^{-i\tilde{k}^ax_a}}{\sqrt{2\tilde{k}^0 V}},
\end{equation}
again with $\tilde{k}^a\tilde{k}_a=0$ and $\tilde{k}^0>0$. Then we have
\begin{equation}
\Delta P_{a}=\frac{\zeta}{2}\int_V \text{d}^{3}\text{x}\left(k_a\tilde{k}_{0}+k_{0}\tilde{k}_a-\eta_{a0}k^c\tilde{k}_c\right)\frac{e^{i\left(k^d-\tilde{k}^d\right)x_d}}{V \sqrt{k^0\tilde{k}^0}}.
\end{equation}
The integration over the spatial coordinates $\mathbf{x}$ leads to a Kronecker delta $\delta_{\mathbf{k},\tilde{\mathbf{k}}}$.
Since both 4-vectors $k^a$ and $\tilde{k}^a$ are null and of positive energy, $\mathbf{k}=\tilde{\mathbf{k}}$ implies that $\tilde{k}^a=k^a$. Then, we obtain
%
\begin{equation}
\Delta P_{a}=\zeta \delta_{\mathbf{k},\tilde{\mathbf{k}}}k_{a}.
\end{equation}
Given a general solution for the field 
$\varphi=\sum (a_\mathbf{k}\varphi_\mathbf{k} + a_\mathbf{k}^*\varphi_\mathbf{k}^*)$ the variation associated with the previous symmetry will result in
\begin{equation}
\Delta P_{a}=\zeta a_{\tilde{\mathbf{k}}} \tilde{k}_{a}.
\end{equation}
One could say that the symmetry extracts from the field its specific Fourier component.
Hence the conserved charges associated with the stress-energy tensor are not invariant under the local symmetries \eqref{eq:emesym1}. This is of course very sensible for a physical symmetry. We have already discussed in Sec. \ref{sec:example} that local symmetries \eqref{eq:emesym1} are not gauge but represent indeed physical symmetries. The transformation of the conserved charges associated with the stress-energy tensor is just another manifestation of this feature: physical symmetries connect different physical solutions which in general have different conserved charges. The stress-energy tensor contains terms which depend directly on the functional form of the free scalar field $\varphi$. When we choose $\varphi \neq 0$ we are kind of imposing a background field (physically there seems to be no reason to justify a particular choice) and different states related by the residual symmetry interact with this background in different solutions, thus having different energies and momenta.

There are two exceptions to this behavior. One of them is the trivial one: for $\zeta=0$ we recover the standard electromagnetic case in which the local transformations \eqref{eq:emesym1} are gauge symmetries, and hence do not change the conserved charges used to parametrize the space of solutions. Most importantly, this is also the case when we consider the projection $\Pi$ to the subset $\varphi=0$ of the space of classical solutions and the local transformations \eqref{eq:emesym1} are understood as emergent gauge symmetries. In this situation, not only the charges are invariant, but the very stress-energy tensor \eqref{eq:SETzeta} is. That is, starting from a stress-energy tensor that is not invariant under a set of local symmetries, which is not surprising due to the physical nature of these symmetries, we end up with a stress-energy tensor which is invariant under the emergent gauge symmetries. From the perspective taken here, this is a peculiarity that stems from the structure of the example analyzed here. For instance, it is reasonable to imagine a situation in which the stress-energy tensor results to be non invariant under the emergent gauge symmetries, but the charges are. In other words, what is observable or not in the effective theory with emergent gauge symmetries depends on the nature of the projection $\Pi$ and the very structure of the theory being analized; it is an outcome, not an a priori imposition. This is of clear importance to the emergent gravity program, as we describe in Sec. \ref{sec:dis}.

\subsection{Emergence of electromagnetism from a Galilean vector field} \label{sec:galilean}

Except for this subsection, in this article we have assumed that a full Lorentz symmetry is at work from the start. Instead, here we are going to discuss how the emergence of a Lorentz symmetry in a Galilean system with only a basic Galilean vector field can lead directly to standard electromagnetism. This subsection is a bit aside the main theme of this article, and can be skipped without loosing the main track. However, we have included it because it resonates with the key ingredients of the rest of the article.

Lorentz symmetry might be an emergent feature of high-energy physics not having this symmetry. Not knowing the characteristics of this high-energy physics, one typical choice is to model it as a Galilean theory, which can be argued to be the simplest space-time structure. Indeed, it is by now well known that it is quite easy and ubiquitous to obtain emergent Lorentz symmetries in Galilean systems, through the presence of Fermi points~\cite{Volovik2008,Volovik2009}. In~\cite{Volovik2009,Liberati2001,Barcelo2007} it is shown how the use of Lorentzian notions might be just an epistemological election, appropriate for the description of a system by internal observers that do not want to use fiduciary external extructures in their description. For example in~\cite{Barcelo2007} it was shown that one can obtain the phenomenology of special relativity, for example the Michelson-Morley experiment, within an emergent scenario of low-energy fields living in a fiduciary external Galilean world. For that a crucial ingredient is that the clocks and rulers used by internal observers in the system are made themselves of the very same low-energy fields.

If the effective low-energy physics of a Galilean theory is described by a single scalar field, we can trivially say that a Lorentz symmetry has emerged if one shows that at low energies this scalar field happens to satisfy a Klein-Gordon equation. In terms of Galilean physics the next spatial structure one could be interested in, thinking in the next level of complexity, is a vector field $V^i$. However, when the relevant effective field in the Galilean world is a vector field, it is not so direct to discern whether some relativistic behavior has emerged.

Imagine that we derive that at low energies the vector field satisfies
\begin{equation}
(\partial_a \partial^a-m^2) V^i = j^i_{\rm P},
\label{eq:eqforV_i}
\end{equation}
where $j^i_{\rm P}$ is some physical source current; for simplicity we are using Cartesian coordinates in the Galilean world.
Being $V^i$ a spatial vector this is not quite a completely covariant relativistic equation. However, some formal manipulations can make the system to look like completely relativistic. One can define a scalar potential with the same dimensions than $V^i$ implicitly through the equation
\begin{equation}
\partial_0 V^0=-\partial_i V^i.
\label{eq:defV_0}
\end{equation}
This is not an independent quantity but completely fixed once one knows $V^i$. 
Now, by assuming that the physical current satisfies a continuity equation one can 
straightforwardly show that
\begin{equation}
\partial_0 (\partial_a \partial^a -m^2) V^0=-\partial_i j^i_{\rm P}= \partial_0 \rho.
\label{eq:consistency1}
\end{equation}
The continuity equation fixes the functional form of the physical density except for the addition of an arbitrary function $f(\mathbf{x})$. Once we choose a specific $\rho_{\rm P}$ we can write
\begin{equation}
(\partial_a \partial^a -m^2) V^0= \rho+f(\mathbf{x}) \equiv \rho_{\rm P}.
\label{eq:consistency2}
\end{equation}
The physical density can be defined precisely as the one choosen to appear as the source of $V^0$.  
The previous equation can be combined with (\ref{eq:eqforV_i}) resulting in 
\begin{equation}
(\partial_a \partial^a -m^2) V^a= j^a_{\rm P},
\label{eq:eqforV_a}
\end{equation}
which is a fully relativistic equation. Once the full Lorentz symmetry is obtained it is a simple matter to decide working in any system of coordinates, not just Cartesian.

Equation \eqref{eq:eqforV_a} is equivalent to the standard Proca equation once we take into account the existence of the Lorenz condition $\partial_a V^a=0$ [which now comes from the definition (\ref{eq:defV_0})]. In the source-free case, they exhibit the symmetry $V_a \rightarrow V_a + \partial_a \chi$ with $(\partial_a \partial^a -m^2)\chi=0$. Moreover, in the case in which one assumes that the mass term can be neglected, these equations are exactly the standard electromagnetic equations in the Lorenz gauge.

Up to here the line of thought we have followed to add the field $V^0$ has been driven by a formal reasoning. However, it is interesting to realize that there are also physical reasons to do that. In a theory with just $V^i$ and equation (\ref{eq:eqforV_i}) one can imagine situations in which $j_{\rm P}^i$ is zero from a specific moment of time on. Imagine for example a distribution of electric charges that become static at that time; let us also assume that there are no free waves present and that the total charge is zero. Then, it is reasonable to think that $V^i$ should be zero from that moment on because of the equation $(\partial_a \partial^a -m^2)V^i=0$. In turn, one could tend to think that the configuration has no energy as the only object available to define an energy is $V^i$. However, thinking for instance in the previous charge configuration, one might start to suspect that this could or could be not the end of the story. 

The definition of $V^0$ as the time integral of $\partial_i V^i$ makes it a variable that {\em knows} how an actual static configuration comes into being from a hypothetical initial state with no sources of any kind. The variable $V^0$ compiles this information; for instance, $V^i$ can be zero from some time on but this does not imply $V^0=0$, only that it does not depend on time. Therefore, the energy of the configuration could depend on the variable $V^0$ and could in general be non-zero even with $V^i=0$. This observation highlights the special status that a variable $V^0$ might have in the theory. In the case in which $V^0$ contributes to the energy of the system this variable acquires a real and independent status as a configuration variable: changes in $V^0$ (time component of the photons) are directly related and only related with changes in $\partial_i V^i$ (longitudinal component of the photons); instead, transverse changes are decoupled from $V^0$ (tranverse components of the photons) and can exist by themselves and propagate freely. 

The situation we end up with is the following. From an emergent perspective, one could identify physical quanties underneath an effective low-energy electromagnetic field. As Maxwell himself did with his mechanical model of electromagnetism (see the discusion in~\cite{Barcelo2014n} and references therein), one could identify some physical quantities playing the role of magnetic and electric fields. However, there is an even simpler description for the electromagnetic phenomena in terms of just one Galilean vector field. Physically this vector field represents some sort of flow field adapted to the flow of charges. Remarkably, a crucial ingredient for this description to be possible is that the instantaneous value of the field $V^i$ happens to be not enough to describe the instantaneous physical configuration. It might happen that to have at an instant all the information about the physical configuration we need also $V^0$, which takes into account how a particular configuration was obtained from the past evolution of the flow fields (i.e. $V^i$ and its relative $j^i_{\rm P}$).
For instance, the stress-energy densities of a theory exhibiting a complete Lorentz symmetry would depend on the vector field $V^i$ and the scalar field $V^0$, the latter one entering in the energy with a negative sign. Only when the theory for a Galilean vector field is of this sort one would have that in the massless case there is no observable that distinguishes a configuration $V^a$ from one ${V'}_a=V_a + \partial_a \chi$. Formally, the system would exhibit a gauge symmetry.

The previous discussion is true as long as the Lorenz condition Eq. \eqref{eq:defV_0} is satisfied. However, there is a degeneracy in this very definition, in the following sense. Eq. \eqref{eq:defV_0} is not the most general equation leading to the Lorentz-invariant field equations \eqref{eq:eqforV_a}. Indeed, the most general implicit definition of $V^0$ leading to the field equations \eqref{eq:eqforV_a} is given by
\begin{equation}
\partial_0V^0=-\partial_i V^i+\varphi,\qquad (\partial_a \partial^a-m^2)\varphi=0.\label{eq:gendefV_0}
\end{equation}
Hence non-trivial solutions for $\varphi$ break the tight connection between the time and longitudinal component of photons. Interestingly, it is for $\varphi=0$ when we recover both this connection and the emergence of gauge symmetries in the massless case $m=0$. Indeed, this setting is essentially the same we have been discussing above for the particular value $\zeta=-1$ (the so-called Feynman gauge). This example leads to additional physical motivation for the imposition of the Lorenz condition, being the most natural among the choices in Eq. \eqref{eq:gendefV_0}. 

To end up this section let us just say that it is remarkable how intertwined is the emergence of a Lorentz symmetry in a Galilean world with only one vector field with the emergence of an effective standard electromagnetism with its gauge symmetry. At the same time it is also remarkable how simple and natural is the scenario just described.

\section{Further discussion and scope} \label{sec:dis}

In the example analyzed in Sec. \ref{sec:example}, starting from a stress-energy tensor that is not invariant under a set of local symmetries we end up with a stress-energy tensor which is invariant under the emergent gauge symmetries. From the perspective taken in this paper this is a peculiarity of the example analyzed, but not a generic feature applicable to an arbitrary system. The emergence of gauge symmetries following the discussion in Sec. \ref{sec:mech} implies that certain quantities lose their operational meaning. The determination of the quantities that retain their operational meaning has to he performed in a case-by-case basis.

This is of relevance to understand the interplay between, e.g., the Weinberg-Witten (second) theorem \cite{Weinberg1980} and the emergent gravity program. This theorem is formulated in the framework of linearized gravity (i.e., described by means the Fierz-Pauli theory of a tensor field $h^{ab}$) and asserts that one cannot construct a non-trivial Lorentz covariant stress-energy tensor for massless spin-two particles or, equivalently, for the tensor field $h^{ab}$. This is true independently of whether these particles are fundamental or composite (see also the discussion in e.g.~\cite{Loebbert2008}). By a Lorentz covariant stress-energy tensor they meant an object with a well-defined notion of local stress-energy densities, that is, an object which is invariant under the gauge transformations of the theory. Assuming that condensed-matter-like systems (or any system in general that does not contain gauge symmetries) do have a well-defined notion of local energy for all their degrees of freedom, one is pushed to conclude that from these degrees of freedom one would not be able to build up something mimicking a spin-two particle. The discussion in this paper invites to reconsider this line of thought. Indeed, it is possible that there exists a way to embed the Fierz-Pauli theory in a theory with no gauge symmetries, and hence with a local notion of energy, so that the former arises due to the projection to a suitable subset of classical solutions, in complete parallelism with our discussion of the electromagnetic example developed in Sec \ref{sec:example}. Note that Fierz-Pauli theory is also a linear theory, which makes the comparison with the electromagnetic example quite acceptable. The only novelty in this case would be that the stress-energy tensor of the (linearized) gravitational field turns out to be non-invariant under the emergent gauge symmetries.

A similar comment applies to the results raised out recently by Marolf \cite{Marolf2015}. He argues that any condensed-matter-like system aiming to reproduce general relativity at low energies should contain some kinematic non-locality. In more general terms, the degrees of freedom of the bulk have to be trivial, leaving only the boundary degrees of freedom. Note that Marolf's results are intrinsically nonlinear and hence the interplay between the particular example developed in this paper and these results is subtler. But again, the suggestion coming from our general discussion in Sec. \ref{sec:mech} is the same. Marolf's results arise from the assumption that the gauge invariance of general relativity has to be preserved in some sense in the condensed-matter-like (or of any other kind) system. We have shown nevertheless in this paper that emergent gauge symmetries can arise from physical symmetries when a natural restriction on the space of solutions of a given theory with only physical symmetries is considered. Hence this opens the possibility to the existence of a theory with no gauge symmetries and with both bulk and boundary degrees of freedom, in which the degrees of freedom of the bulk are frozen due to the existence of a natural (local) projection to a subset of the space of classical solutions, and not as a characteristic of the very theory. Of course, at the moment this is only an intriguing possibility; we are currently studying the construction of a specific example that materializes this possibility. This is inextricably related to the point raised in the paragraph above concerning the Weinberg-Witten theorem. A quite natural possibility we are considering is starting with a linear theory leading to a satisfactory picture of emergence of gauge symmetries in Fierz-Pauli theory along the lines sketched in the paragraph above, and then complete this construction in a nonlinear fashion following the same rules that permit to obtain general relativity from Fierz-Pauli theory through the inclusion of self-interactions \cite{Deser1970,Deser2010,Carballo2014}. Results of this ongoing research will be reported elsewhere.

\section*{Acknowledgments}

The authors want to thank V\'ictor Aldaya for enlightling conversations related to symmetries in dynamical systems. FDF wants to thanks Gaetano Lambiase for numerous discussions and the Instituto de Astrof\'isica de Andaluc\'ia, IAA - CSIC for its warm hospitality during a visit while most of this work has been done; also support offered by the EU through the Erasmus program. RCR acknowledges support from the Claude Leon Foundation. Financial support was provided by the Spanish MINECO through the projects FIS2014-54800-C2-1, FIS2014-54800-C2-2 (with FEDER contribution), and by the Junta de Andalucía through the project FQM219. 


\bibliographystyle{ieeetr}
\bibliography{references-9}

\begin{thebibliography}{10}

\bibitem{Carlip2001}
S.~Carlip, ``{Quantum gravity: A Progress report},'' {\em Rept. Prog. Phys.},
  vol.~64, p.~885, 2001.

\bibitem{Thiemann2007}
T.~Thiemann, {\em {Modern canonical quantum general relativity}}.
\newblock Cambridge University Press, 2008.

\bibitem{Carlip2012}
S.~Carlip, ``{Challenges for Emergent Gravity},'' {\em Stud. Hist. Philos. Mod.
  Phys.}, vol.~46, pp.~200--208, 2014.

\bibitem{Weinberg1980}
S.~{Weinberg} and E.~{Witten}, ``{Limits on massless particles},'' {\em Phys.
  Lett. B}, vol.~96, pp.~59--62, 1980.

\bibitem{Marolf2015}
D.~Marolf, ``{Emergent Gravity Requires Kinematic Nonlocality},'' {\em Phys.
  Rev. Lett.}, vol.~114, no.~3, p.~031104, 2015.

\bibitem{Baskaran1988}
G.~Baskaran and P.~W. Anderson, ``Gauge theory of high-temperature
  superconductors and strongly correlated fermi systems,'' {\em Phys. Rev. B},
  vol.~37, pp.~580--583, Jan 1988.

\bibitem{HenneauxTeitelboim1992}
M.~Henneaux and C.~Teitelboim, {\em {Quantization of gauge systems}}.
\newblock 1992.

\bibitem{MisnerThorneWheeler1974}
C.~W. Misner, K.~S. Thorne, and J.~A. Wheeler, {\em {Gravitation}}.
\newblock San Francisco: W. H. Freeman, 1973.

\bibitem{Banados2016}
M.~Ba\~nados and I.~A. Reyes, ``{A short review on Noether's theorems, gauge
  symmetries and boundary terms},'' 2016.

\bibitem{Wipf1993}
A.~W. Wipf, ``{Hamilton's formalism for systems with constraints},'' 1993.
\newblock [Lect. Notes Phys.434,22(1994)].

\bibitem{Julia1998}
B.~Julia and S.~Silva, ``{Currents and superpotentials in classical gauge
  invariant theories. 1. Local results with applications to perfect fluids and
  general relativity},'' {\em Class. Quant. Grav.}, vol.~15, pp.~2173--2215,
  1998.

\bibitem{Aldaya2010}
V.~Aldaya, M.~Calixto, J.~Guerrero, and F.~F. Lopez-Ruiz, ``{Symmetries of
  Non-Linear Systems: Group Approach to their Quantization},'' {\em Int. J.
  Geom. Meth. Mod. Phys.}, vol.~8, pp.~1329--1354, 2011.

\bibitem{Barcelo2014n}
C.~{Barcel{\'o}}, R.~{Carballo-Rubio}, L.~J. {Garay}, and G.~{Jannes},
  ``{Electromagnetism as an emergent phenomenon: a step-by-step guide},'' {\em
  New J. Phys.}, vol.~16, no.~12, p.~123028, 2014.

\bibitem{Itzykson2012}
C.~Itzykson and J.~B. Zuber, {\em Quantum Field Theory}.
\newblock Dover Books on Physics, Dover Publications, 2012.

\bibitem{Jimenez2008}
J.~Beltr\'{a}n~Jim\'{e}nez and A.~L. Maroto, ``{Cosmological electromagnetic
  fields and dark energy},'' {\em JCAP}, vol.~0903, p.~016, 2009.

\bibitem{Jimenez2009}
J.~Beltr\'{a}n~Jim\'{e}nez and A.~L. Maroto, ``{The electromagnetic dark
  sector},'' {\em Phys. Lett.}, vol.~B686, pp.~175--180, 2010.

\bibitem{Jimenez2009b}
J.~Beltr\'{a}n~Jim\'{e}nez and A.~L. Maroto, ``{Dark energy: The Absolute
  electric potential of the universe},'' {\em Int. J. Mod. Phys.}, vol.~D18,
  pp.~2243--2248, 2009.

\bibitem{Maroto2011}
J.~{Beltr\'{a}n Jim{\'e}nez} and A.~L. {Maroto}, ``{The Dark Magnetism of the
  Universe},'' {\em Mod. Phys. Lett. A}, vol.~26, pp.~3025--3039, 2011.

\bibitem{Weinberg1972}
S.~Weinberg, {\em {Gravitation and Cosmology: principles and applications of
  the general theory of relativity}}.
\newblock New York: John Wiley and Sons, 1972.

\bibitem{Volovik2008}
G.~{Volovik}, ``{Emergent physics: Fermi-point scenario},'' {\em Philos. T.
  Roy. Soc. A}, vol.~366, pp.~2935--2951, 2008.

\bibitem{Volovik2009}
G.~E. Volovik, {\em The Universe in a Helium Droplet}.
\newblock International Series of Monographs on Physics, OUP Oxford, 2009.

\bibitem{Liberati2001}
S.~Liberati, S.~Sonego, and M.~Visser, ``{Faster than c signals, special
  relativity, and causality},'' {\em Annals Phys.}, vol.~298, pp.~167--185,
  2002.

\bibitem{Barcelo2007}
C.~Barcel\'{o} and G.~Jannes, ``{A Real Lorentz-FitzGerald contraction},'' {\em
  Found. Phys.}, vol.~38, pp.~191--199, 2008.

\bibitem{Loebbert2008}
F.~Loebbert, ``{The Weinberg-Witten theorem on massless particles: An Essay},''
  {\em Ann. Phys.}, vol.~17, pp.~803--829, 2008.

\bibitem{Deser1970}
S.~{Deser}, ``{Self-interaction and gauge invariance},'' {\em Gen. Relat.
  Gravit.}, vol.~1, pp.~9--18, 1970.

\bibitem{Deser2010}
S.~{Deser}, ``{Gravity from self-interaction redux},'' {\em Gen. Relat.
  Gravit.}, vol.~42, pp.~641--646, 2010.

\bibitem{Carballo2014}
C.~Barcel\'o, R.~Carballo-Rubio, and L.~J. Garay, ``Unimodular gravity and
  general relativity from graviton self-interactions,'' {\em Phys. Rev. D},
  vol.~89, p.~124019, Jun 2014.

\end{thebibliography}

\end{document}